\documentclass[aps,pra,twocolumn,showpacs,preprintnumbers,amsmath,amssymb,floatfix,superscriptaddress]{revtex4-1}

\usepackage{graphics}
\usepackage{graphicx}
\usepackage{amsmath}
\usepackage{setspace}
\usepackage{braket}
\usepackage{epstopdf}
\usepackage{comment}
\usepackage{hyperref}
\usepackage{bm}
\usepackage{xfrac}
\usepackage{xcolor}

\usepackage{mathrsfs}

\hypersetup{%
   pdfpagemode=None, 
   pdfstartpage=1,
   pdfmenubar=true,
   pdftoolbar=true,
   colorlinks = true,
   linkcolor=blue,
   citecolor=blue,
   urlcolor=blue,
   bookmarksopen=false
 }

\newcommand{\tj}[6]{ \begin{pmatrix}
  #1 & #2 & #3 \\
  #4 & #5 & #6 
 \end{pmatrix}}

\begin{document}
\author{Jes\'us P\'{e}rez-R\'{i}os}
\address{Fritz-Haber-Institut der Max-Planck-Gesellschaft, Faradayweg 4-6, D-14195 Berlin, Germany}
\address{Institute for Molecules and Materials, Radboud University, Heyendaalseweg 135, 6525 AJ Nijmegen, The Netherlands}

\title{Electric field dissociation of weakly bound molecular ions}

\date{\today}

\begin{abstract}
 
We present a full quantal study on the dissociation of a weakly bound molecular ion in the presence of an external time-dependent electric field. We focus on the dissociation dynamics of a molecular ion in a Paul trap relevant for atom-ion hybrid traps. Our results show that a weakly bound molecular ion survives in a Paul trap giving a theoretical ground to previous experimental findings [A. Kr\"ukow et al. Phys. Rev. Lett. 116, 193201 (2016) and A. Mohammadi et al. Phys. Rev. Research 3, 013196 (2021)]. In particular, we find that weakly bound molecular ions are more likely to survive in traps with large RF frequency. Similarly, we show that applying an electric field ramp is an efficient method to state-selectively detect weakly bound molecular ions, analogous to the well-known selective field ionization technique applied in Rydberg atoms, that it may also be used to detect these species in atom-ion hybrid traps.

\end{abstract}

\maketitle

The advent of hybrid neutral-ion traps has boosted cold chemistry research due to the possibility of bringing together ions and atoms in a controlled manner~\cite{JPRBook,COTE20166,RevModPhys.91.035001,Julienne2012,Hudson2019}. Similarly, these traps find applications in different research areas such as the development of new and more efficient quantum information protocols~\cite{QC1,QC2,QC3,QC4,QC5,QC6,QC7,QC8}, the realization of quantum logic spectroscopy schemes~\cite{Mur2012,QLSchapter,Wolf2016,Sinhal12020,Najafian20202,Collopy2021,PhysRevLett.125.120501} and the study of impurity physics~\cite{Kleinbach_2018,Meso1,Meso2,Meso3,astrakharchik2020ionic,Massignan2005,dieterle2020transport,Lemeshko2016,Mukherjee2015}, to cite a few. On the impurity physics front, when a single charged impurity, A$^+$, is brought in contact with an ultracold atomic gas B, at sufficient densities, the ion undergoes a three-body recombination reaction: A$^+$ + B + B$\rightarrow$ AB$^+$ + B leading to the formation of weakly bound molecular ions\cite{Artjom2016,JPRBook,JPR2021}. In principle, these molecular ions would be hard to observe since the oscillating electric field of a Paul trap could tear them apart, or the presence of external laser sources could photodissociate them~\cite{Amir2021}. However, these weakly bound molecular ions are experimentally observed, and with it, the validity and accuracy of the classical trajectory method to calculate the three-body recombination rate is corroborated~\cite{Artjom2016,dieterle2020transport,Amir2021}. After its formation, these molecular ions relax very efficiently into deeper bound vibrational states via collisions with ultracold atoms\cite{JPR2019,Jachymski2020} opening up photodissociation routes through the external laser sources characteristic of atom-ion hybrid traps~\cite{Amir2021}.

Similarly, it has been recently predicted that a single ion in an ultracold molecular bath reacts and forms weakly bound molecular ions~\cite{Hirzler2020}, and the same seems to occur in the case of a microwave ion clock when a single ion reacts with background gases~\cite{YbH+,YbH+2,YbH+3}. In the case of a single ion in a molecular bath the simulations indicate that molecular ions survive despite the time-dependent electric field of the Paul trap~\cite{Hirzler2020}. However, the complex reaction network of a single ion in an atomic or molecular bath and the need of a quantal treatment may raise legit questions about the net effect of the time-varying electric fields on the molecular ions. Indeed, there has been little theoretical effort towards understanding the dynamics of a weakly bound molecular ion in the presence of an electric field, barring a static or effective model approach~\cite{dieterle2020transport}. 

In this letter, we present a full quantum study of the dissociation dynamics of a weakly bound molecular ion in the presence of a time-dependent electric field. In particular, we mimic the electric field that an ion feels in a Paul trap characteristic of atom-ion hybrid traps, finding that the survival of a molecular ion depends on its binding energy and the trap frequency. In addition, we consider a Gaussian-type pulse to simulate an electric field ramp meant to dissociate the molecular ion in a controllable manner. As a result, we show that it is possible to probe the vibrational states of a weakly bound molecular ion individually via the electric field ramp method similar to the state selective field ionization method used to ionize Rydberg atoms.

\begin{figure*}
    \center
    \includegraphics[width=\linewidth]{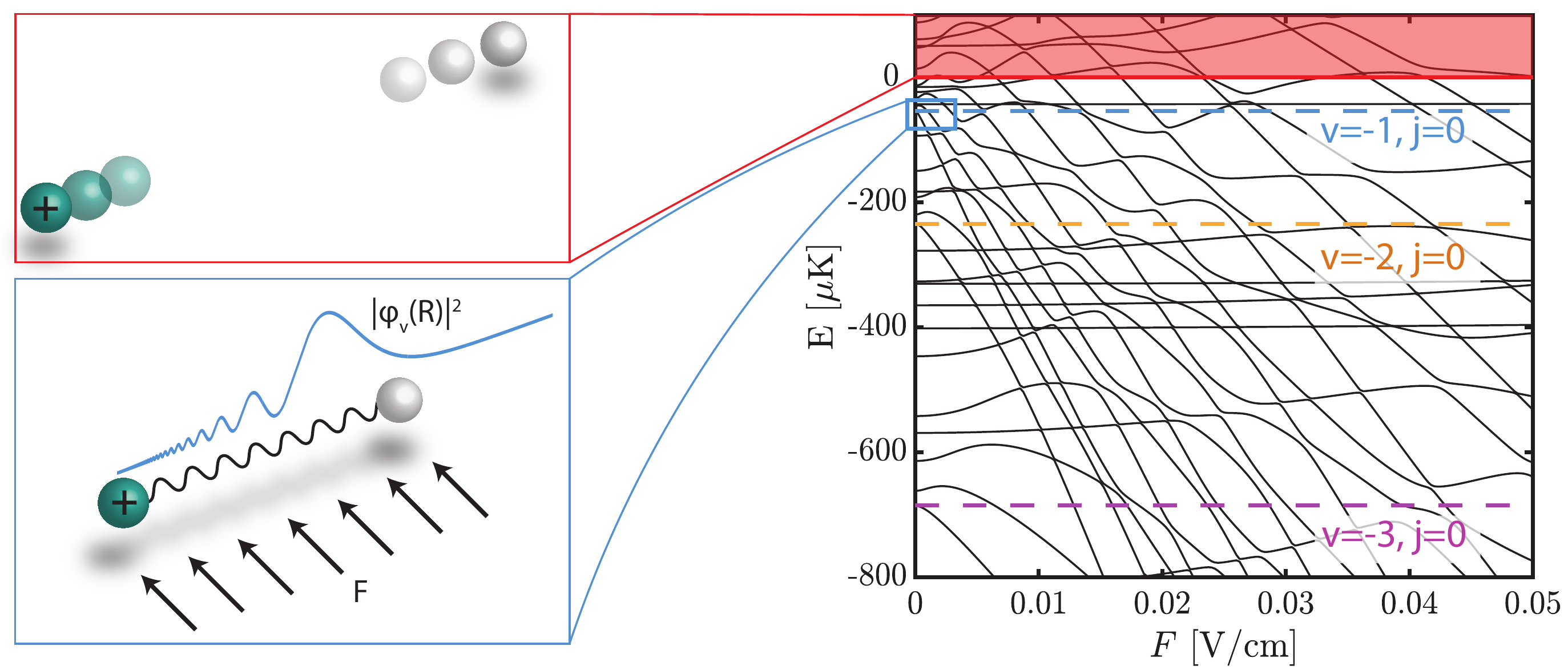}
    \caption{Rovibrational energy spectrum of the  $^{174}$Yb$^{87}$Rb$^{+}$ molecular ion as a function of the electric field strength $F$. The dashed lines corresponds to a few of the relevant rovibrational states labeled by the vibratioanal quantum number $v$ and the rotational quantum number $j$. The shaded region represents the onset of dissociation. The left panel displays schematically the nature of the molecular ion at different electric field strengths; the lower panel represents a weakly bound molecular ion whereas the upper panel stands for a dissociated molecular ion.}
    \label{fig1}
\end{figure*}

The dynamics of a molecular ion in an external electric field can be described by the following Hamiltonian

\begin{equation}
\label{eq1}
    H=H_0+H_{\text{int}}(t),
\end{equation}

\noindent
where $H_0$ represents the unperturbed part of the rovibrational Hamiltonian of the molecular ion. Therefore, $H_0$ is diagonal in the basis of the vibrational $v$ and rotational $j$ states, i.e., $H_0|vjm\rangle=E_{vj}|vjm\rangle$, where $E_{vj}=E_v+B_v\hbar^2j(j+1)$, $\hbar$ being the reduced Planck constant, $E_v$ is the vibrational energy of the $v$-th rotational state, $B_v$ is the rotational constant of the assumed rigid rotor in the same vibrational state, and $m$ is the projection of the rotational quantum number on the quantization axis. $E_v$ and $B_v$ are calculated assuming a given atom-ion interaction. In Eq.~(\ref{eq1}) the time-dependent part of the Hamiltonian is given by the interaction between the dipole moment of the molecular ion $\vec{d}$ and the time-dependent external electric field $F(t)$ as

\begin{equation}
\label{eq2}
   H_{\text{int}}(t)= -\vec{d} \cdot\vec{F}(t),
\end{equation}

\noindent
where we assume $\vec{d}\equiv q\vec{R}$, with $q$ denoting the charge of the ion and $R$ representing the relative position vector between the atoms within the diatomic molecular ion. 

First, we calculate eigenvalues of the Hamiltonian (\ref{eq1}) for $^{174}$Yb$^{87}$Rb$^{+}$ with $m=0$, as an example, as a function of the electric field strength, including ten vibrational states and 20 rotational states with $j\le 19$ ensuring the convergence of the results. It is worth noticing that the vibrational states are labeled from the dissociation threshold in this work, i.e., $v=-1$ is the vibrational state with the smallest binding energy, and the larger the value of $v$, the larger the binding energy. The atom-ion interaction is taken as $-C_4/R^{4}\left(1-\frac{1}{2}\left(\frac{R_m}{R} \right)^4\right)$ with $C_4=160$~a.u. and $R_m$=10.142~a$_0$ (a$_0\equiv0.529177\times 10^{-10}$~m) is the equilibrium distance, corresponding to the a$^3\Sigma^{+}$ electronic state of the system~\cite{PotYbRbplus}. The vibrational wave functions are obtained by numerically solving the time-independent Schr\"odinger equation for the vibrational motion using the Numerov method with an equispaced radial grid of $10^5$ steps between 5.25~a$_0$ and 1725~a$_0$ using $10^5$ steps to ensure a convergence of the vibrational energies within 0.1$\%$. 

In Fig.~\ref{fig1} we show the rovibrational energy spectrum of the least bound states of the $^{174}$Yb$^{87}$Rb$^{+}$ molecular ion as a function of the electric field strength, where it is noticed that the pathway towards dissociation (the red shaded region) is somewhat complex and highly dependent on the initial state. However, we find that the $j=0$ states lower their energy as the electric field magnitude increases, whereas high $j$ states show the opposite behavior. Therefore, $j=0$ states will require a larger electric field to dissociate than states with higher $j$ values. In the same vein, focusing on the $|v=-1,j=0\rangle$ state, we observe that it interacts with many close rovibrational states at electric fields $\lesssim$~0.01 V/cm (the same order of magnitude as the electric field that a molecular ion feels in a Paul trap)  whereas higher $j$ states of the same vibrational state show a more direct pathway into the dissociation limit (the shaded red area). As a result, a static picture in which the dipole-electric field coupling lowers the dissociation threshold leading to the dissociation of the weakly bound states is an oversimplification of the real scenario.

To fully understand the dynamics of a weakly bound molecular ion in a time-dependent field we solve the time-dependent Schr\"odinger equation with the Hamiltonian (\ref{eq1}) via the ansatz

\begin{equation}
\label{eq3}
    |\psi(t)\rangle=\sum_{vjm}c_{vjm}(t)|vjm\rangle e^{-\imath E_{vj}t/\hbar}, 
\end{equation}

\noindent
yielding the following system of coupled first-order differential equations 

\begin{eqnarray}
\frac{dc_{v'j'm'}(t)}{dt}&=&-\frac{\imath}{\hbar}\sum_{vjm}   \langle v'j'm'|H_{\text{int}}(t)|vjm\rangle c_{vjm}(t) \nonumber \\
& &\times e^{-\imath \left( E_{v'j'} - E_{vj}\right) t/\hbar}.
\end{eqnarray}

\noindent
The coupling between different rovibrational states is mediated by the dipole moment of the molecular ion and reads as

\begin{eqnarray}
\label{eq4}
    \langle v'j'm'|H_{\text{int}}(t)|vjm\rangle= 
   -\langle v'|d|v\rangle F(t)\delta_{mm'}(-1)^{m'}\nonumber \\
   \sqrt{(2j+1)(2j'+1)}\tj{j}{1}{j'}{m}{0}{-m} \tj{j}{1}{j'}{0}{0}{0},
\end{eqnarray}
where
\begin{equation}
 \langle v'|d|v\rangle=q\int_0^{\infty}\phi^{*}_{v'}(R)R\phi_{v}(R)dR,
\end{equation}

\noindent
and $(...)$ stands for the 3j symbol. Eq.~(\ref{eq4}) includes, apart from the expected rotational coupling via the 3j symbols, the intra-vibrational coupling through the dipole moment of the molecular ion. 

\begin{figure}
    \center
    \includegraphics[width=\linewidth]{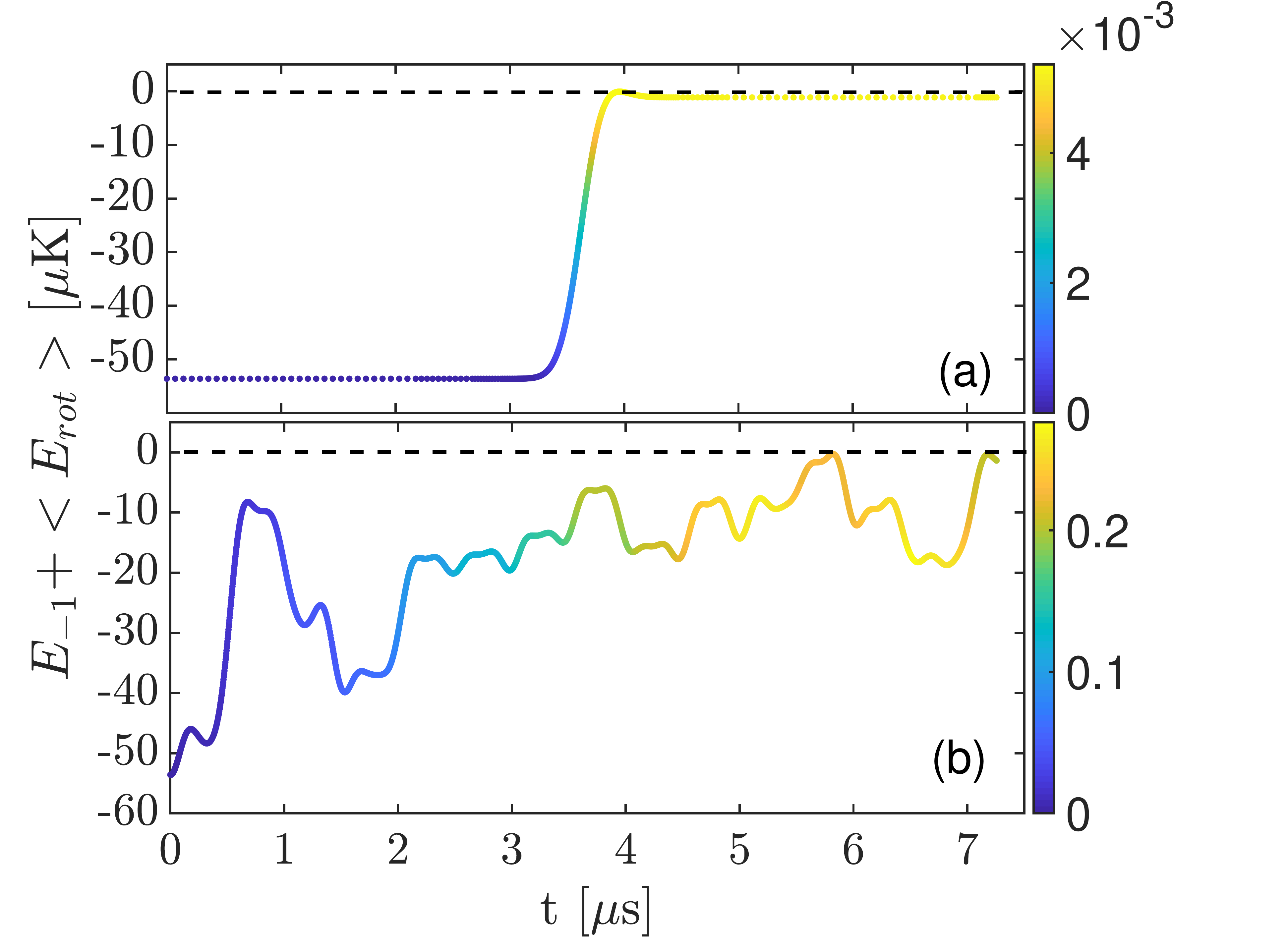}
    \caption{Molecular ion energy of the $(v=-1;j=0)$ state of $^{174}$Yb$^{87}$Rb$^{+}$ as a function of time for different electric field pulses. Panel (a) is for a Gaussian pulse centered at $t_0=\frac{T_{\text{rot}}}{2}=\frac{\pi}{2B_{-1}}$ with $\tau=$~500ns and $F_0=$~0.00168 V/cm whereas in panel (b) refers to RF-type electric field with $\omega_{\text{RF}}=$~$2\pi\times 2$~MHz, $F_{DC}=$~0.001~V/cm and $F_{AC}=$~0.00156~V/cm. The color code refers to the probability of populating rovibrational states with $v< -1$, i.e, the vibrational quenching probability. }
    \label{fig2}
\end{figure}

The time-evolution of the energy associated with the $|v=-1,j=0\rangle$ state of $^{174}$Yb$^{87}$Rb$^{+}$ for different time-dependent electric fields is shown in Fig.~\ref{fig2}. Panel (a) displays the result for an electric field ramp to $F_0$ in a time $\tau$, which is simulated with a Gaussian pulse $F_{\text{G}}(t)=F_0\exp{(-4\ln{(2)}(t-t_0)^2/\tau^2)}$ centered at time $t_0$, with electric field strength $F_0$ and full width half maximum $\tau$~\cite{Bretislav}. Whereas, panel (b) refers to an RF field similar to the one a molecular ion feels in a Paul trap as $F_{\text{RF}}(t)=F_{\text{DC}}+F_{\text{AC}}\cos{(\omega_{RF}t)}$, where $F_{\text{DC}}$ and $F_{\text{AC}}$ represent the DC and AC electric field components of the trap, respectively, and $\omega_{RF}$ stands for the RF trap frequency. The energy presented is the result of adding the time-dependent rotational energy $\langle E_{\text{rot}} \rangle=\sum_{vj}|c_{vj0}(t)|^2B_v\hbar^2j(j+1)$ (where only $m=0$ states are considered) to the initial energy of the molecular ion's state. Indeed, as soon as this magnitude surpasses zero, the molecule dissociates. This criterion for dissociation considers the effect of nearby vibrational states apart from the expected rotational mixing due to the electric field-dipole interaction. As a result, as expected, the time-dependent shape of the electric field plays a major role in the dynamics of the system: for a Gaussian pulse, the molecular ion dissociates at t$\sim\tau$ whereas, for the RF field, it takes a long time depending on the  $\omega_{\text{RF}}$. Another interesting observation from Fig.~\ref{fig2} is to realize that a time-dependent electric field induces a quenching of the vibrational state of the molecular ion. In particular, the vibrational quenching probability is more pronounced in the case of an RF field than in a Gaussian pulse, implying that the dissociation pathway for a Gaussian pulse is more direct than in the case of an RF field.

\begin{figure}
    \center
    \includegraphics[width=\linewidth]{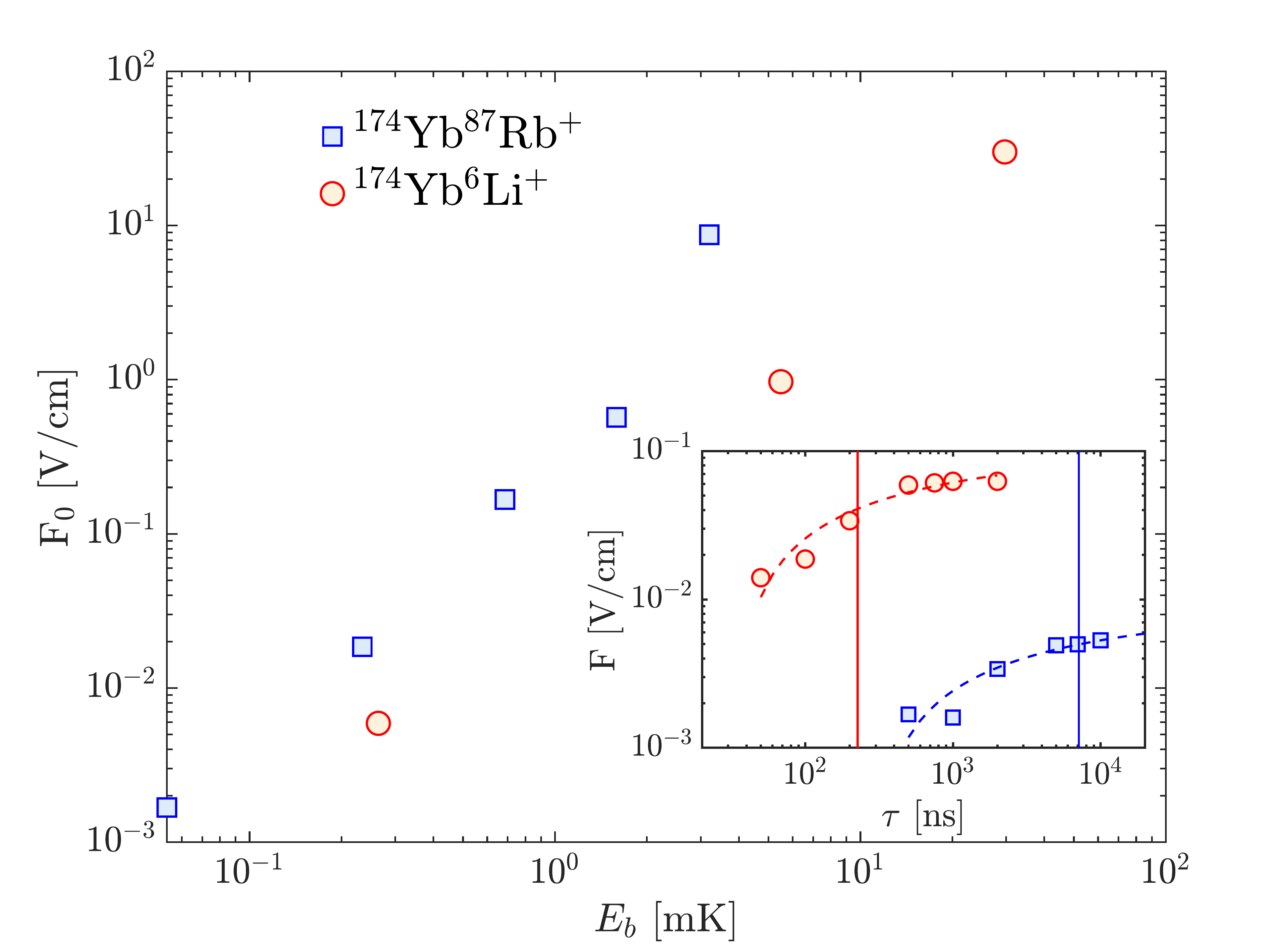}
    \caption{Dissociation field strength for different vibrational states of $^{174}$Yb$^{87}$Rb$^{+}$ and $^{174}$Yb$^{6}$Li$^{+}$ assuming Gaussian pulse with $\tau=$~100ns. The inset shows the electric field dissociation for $v=-1$ states as a function of the pulse duration $\tau$. The dashed lines correspond to a fitting of the data as explained in the text whereas the vertical lines represent the rotational period of the involved vibrational states accordingly with the color code employed in the main figure.}
    \label{fig3}
\end{figure}

A detailed study on the dissociation dynamics of a molecular ion due to a Gaussian pulse is shown in Fig.~\ref{fig3} where the electric field strength for dissociating different vibrational states with $j=0$ are displayed. The calculations for $^{174}$Yb$^{6}$Li$^{+}$ include five vibrational states and 20 rotational states with $j\le 20$ and the corresponding atom-ion interaction is characterized by $R_m=7.59$~a$_0$ and $C_4=82.1$~a.u., which correlates with its a$^3\Sigma^{+}$ electronic state \cite{PotYbLiplus}. We notice that the larger the binding energy the larger the electric field strength needs to be, as expected. The duration of the pulse, $\tau$, affects the dynamics of the the ion on the electric field and the results are shown in the inset of Fig.~\ref{fig3}.For shorter pulses a smaller peak intensity of the electric field is required for dissociation whereas for a duration $\tau\gtrsim T_{\text{rot}}$, where $T_{\text{rot}}=\pi/B_{-1}$ represents the rotational period (it is illustrated as the vertical solid lines), the electric field strength for dissociation tends to a constant value since the system crosses the threshold of the adiabatic limit. Indeed, we notice that the electric field strength for dissociation depends on the duration of the pulse as $F(\tau)=a+b \tau^{-1/4}$, and it is independent of the molecular ion species in question. For the two systems under consideration we find $a=0.009\pm0.001$~V/cm and $b=-0.04\pm0.01$~V/cm$\times$ns$^{1/4}$ for the $|v=-1,j=0\rangle$ state of $^{174}$Yb$^{87}$Rb$^{+}$, whereas $a=0.107\pm 0.007$~V/cm, and $b=-0.26\pm0.03$~V/cm$\times$ns$^{1/4}$ for the $|v=-1,j=0\rangle$ state of $^{174}$Yb$^{6}$Li$^{+}$.

\begin{figure}
    \center
    \includegraphics[width=\linewidth]{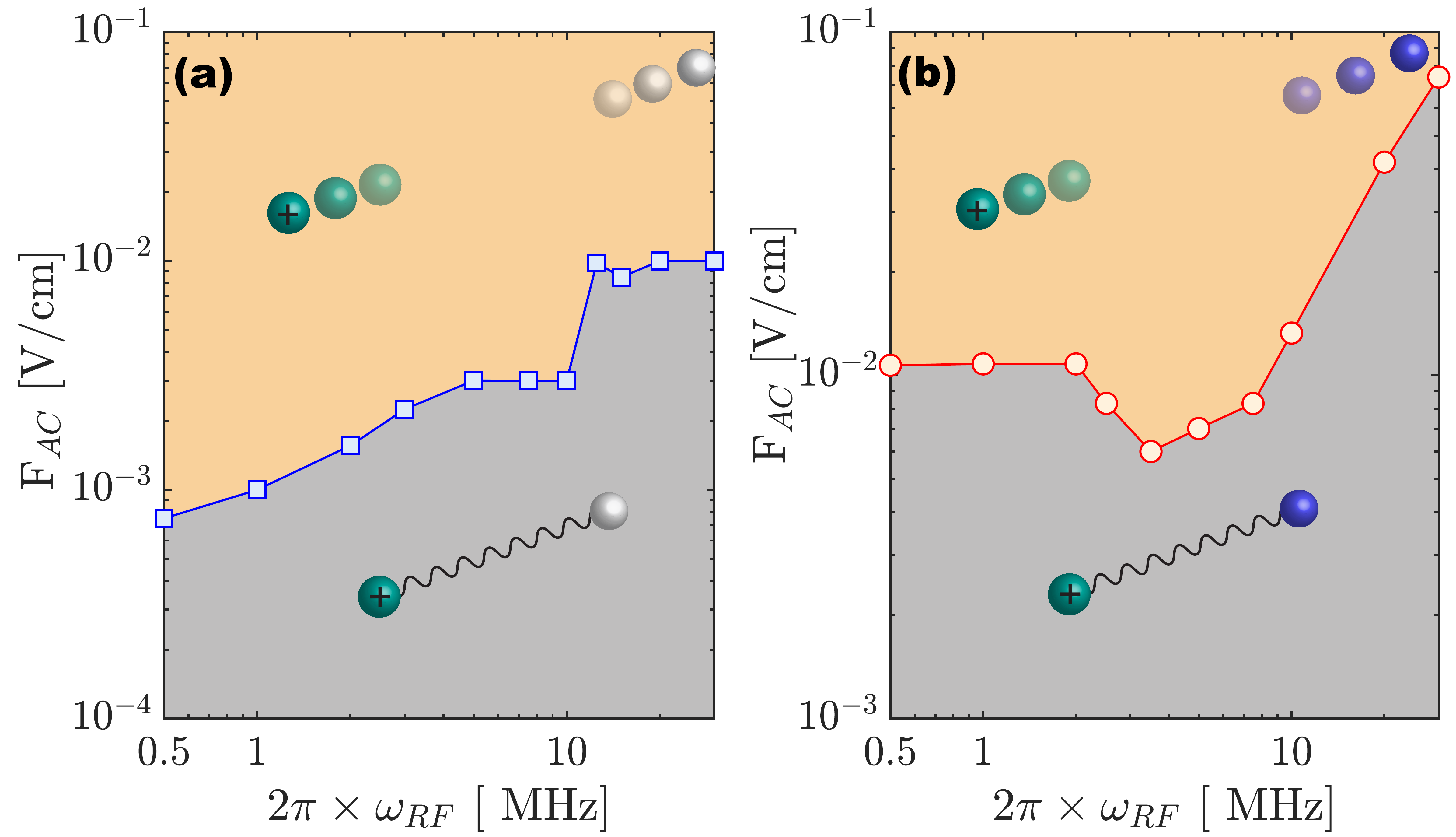}
    \caption{Phase diagram for the dissociation of a weakly bound molecular ion. The phase diagram presents the region of the parameter space for which a given AC electric field $F_{AC}$ dissociates  the molecular ion for a given $\omega_{\text{RF}}$ assuming a fixed DC field value of 0.001 V/cm. Panel (a) refers to the $|v=-1,j=0\rangle$ state for $^{174}$Yb$^{87}$Rb$^{+}$ whereas panel (b) is for $^{174}$Yb$^{6}$Li$^{+}$.}
    \label{fig4}
\end{figure}

The effects of an RF-type electric field over a weakly bound molecular ion are summarized in Fig.~\ref{fig4}, where the maximum amplitude of the applied AC electric field is plotted versus the RF frequency of the trap. As a result, we are able to present a ``phase space'' for dissociation of a weakly bound molecular ion in a Paul trap. The phase space shows two regions, one in which for a given RF frequency the molecular ion survives in a Paul trap and a second one where the molecular ion dissociates as it is schematically presented in the figure. The borderline between those regions depends in an involved manner on the the trap frequency, the mass of the molecular ion and its binding energy. The DC electric field plays an important role on determining the border line between the two phases. In particular, we find that for larger DC electric fields it is necessary a smaller AC field amplitude to dissociate a molecular ion at a given frequency. The typical electric field strengths for dissociation range between 0.001V/cm to 0.1V/cm for RF frequencies between 500kHz and 30MHz. Nevertheless, we identify a general trend: for larger trap frequencies it is necessary to have a larger AC field to dissociate the molecule. Therefore, weakly bound molecular ions are more likely to survive in high frequency traps. 

The magnitude of the AC electric field that a molecular ion feels in a Paul trap depends on the position of the ion from the center of the trap, which in a cold environment it depends mainly on the kinetic energy released after its formation via three-body recombination~\cite{Artjom2016,JPR2015,JPRBook,JPR2021} or ion-molecule collisions~\cite{JPR2019,Hirzler2020,JPR2021}. Indeed, the binding energy of the molecular ion is of the same order of magnitude as the collision energy~\cite{Artjom2016} and in most of the traps $F_{AC}\sim$0.01V/cm. Therefore, $^{174}$Yb$^{6}$Li$^{+}$ after its formation via the reaction Yb$^+$ + Li$_2\rightarrow$ YbLi$^+$ + Li survives in a Paul trap. On the contrary, $^{174}$Yb$^{87}$Rb$^{+}$ dissociates under similar circumstances unless a large trap frequency is used.

In summary, we have shown that the shape of the time-dependent electric field has a drastic effect on the dissociation dynamics of a weakly bound molecular ion. As a result, after defining the phase space for molecular ion dissociation, we find that a weakly bound molecular ion is hardly dissociated in an RF trap unless minimal RF frequencies are employed, and the required electric field depends drastically on the binding energy of the molecular ion. These results confirm previous experimental observations and reinforce theoretical predictions by including the Paul trap on the system's dynamics. Last but not the least, we have shown that an electric field ramp is an effective method to state-selectively detect molecular ions, which may have a potential impact on detecting molecular ion products of relevant cold chemical reactions such as three-body recombination and ion-molecule collisions. Finally, we would like to point out another possible mechanism for molecular ion dissociation: photo-dissociation. In this case, the RF trap frequency is in resonance to a continuum state, which is an idea currently being investigated.

The author would like to thank M. Karra and M. Mirahmadi for valuable suggestions and exciting discussions. Similarly, the insight of H. Hirzler regarding different experimental aspects of this work is highly appreciated.




\begin{thebibliography}{42}%
\makeatletter
\providecommand \@ifxundefined [1]{%
 \@ifx{#1\undefined}
}%
\providecommand \@ifnum [1]{%
 \ifnum #1\expandafter \@firstoftwo
 \else \expandafter \@secondoftwo
 \fi
}%
\providecommand \@ifx [1]{%
 \ifx #1\expandafter \@firstoftwo
 \else \expandafter \@secondoftwo
 \fi
}%
\providecommand \natexlab [1]{#1}%
\providecommand \enquote  [1]{``#1''}%
\providecommand \bibnamefont  [1]{#1}%
\providecommand \bibfnamefont [1]{#1}%
\providecommand \citenamefont [1]{#1}%
\providecommand \href@noop [0]{\@secondoftwo}%
\providecommand \href [0]{\begingroup \@sanitize@url \@href}%
\providecommand \@href[1]{\@@startlink{#1}\@@href}%
\providecommand \@@href[1]{\endgroup#1\@@endlink}%
\providecommand \@sanitize@url [0]{\catcode `\\12\catcode `\$12\catcode
  `\&12\catcode `\#12\catcode `\^12\catcode `\_12\catcode `\%12\relax}%
\providecommand \@@startlink[1]{}%
\providecommand \@@endlink[0]{}%
\providecommand \url  [0]{\begingroup\@sanitize@url \@url }%
\providecommand \@url [1]{\endgroup\@href {#1}{\urlprefix }}%
\providecommand \urlprefix  [0]{URL }%
\providecommand \Eprint [0]{\href }%
\providecommand \doibase [0]{http://dx.doi.org/}%
\providecommand \selectlanguage [0]{\@gobble}%
\providecommand \bibinfo  [0]{\@secondoftwo}%
\providecommand \bibfield  [0]{\@secondoftwo}%
\providecommand \translation [1]{[#1]}%
\providecommand \BibitemOpen [0]{}%
\providecommand \bibitemStop [0]{}%
\providecommand \bibitemNoStop [0]{.\EOS\space}%
\providecommand \EOS [0]{\spacefactor3000\relax}%
\providecommand \BibitemShut  [1]{\csname bibitem#1\endcsname}%
\let\auto@bib@innerbib\@empty
\bibitem [{\citenamefont {R\'ios}(2020)}]{JPRBook}%
  \BibitemOpen
  \bibfield  {author} {\bibinfo {author} {\bibfnamefont {J.~P.}\ \bibnamefont
  {R\'ios}},\ }\href@noop {} {\emph {\bibinfo {title} {An introduction to cold
  and ultracold chemistry}}}\ (\bibinfo  {publisher} {Springer},\ \bibinfo
  {address} {Cham, Switzerland},\ \bibinfo {year} {2020})\BibitemShut {NoStop}%
\bibitem [{\citenamefont {Côté}(2016)}]{COTE20166}%
  \BibitemOpen
  \bibfield  {author} {\bibinfo {author} {\bibfnamefont {R.}~\bibnamefont
  {Côté}}\ }(\bibinfo  {publisher} {Academic Press},\ \bibinfo {year}
  {2016})\ pp.\ \bibinfo {pages} {67 -- 126}\BibitemShut {NoStop}%
\bibitem [{\citenamefont {Tomza}\ \emph {et~al.}(2019)\citenamefont {Tomza},
  \citenamefont {Jachymski}, \citenamefont {Gerritsma}, \citenamefont
  {Negretti}, \citenamefont {Calarco}, \citenamefont {Idziaszek},\ and\
  \citenamefont {Julienne}}]{RevModPhys.91.035001}%
  \BibitemOpen
  \bibfield  {author} {\bibinfo {author} {\bibfnamefont {M.}~\bibnamefont
  {Tomza}}, \bibinfo {author} {\bibfnamefont {K.}~\bibnamefont {Jachymski}},
  \bibinfo {author} {\bibfnamefont {R.}~\bibnamefont {Gerritsma}}, \bibinfo
  {author} {\bibfnamefont {A.}~\bibnamefont {Negretti}}, \bibinfo {author}
  {\bibfnamefont {T.}~\bibnamefont {Calarco}}, \bibinfo {author} {\bibfnamefont
  {Z.}~\bibnamefont {Idziaszek}}, \ and\ \bibinfo {author} {\bibfnamefont
  {P.~S.}\ \bibnamefont {Julienne}},\ }\href {\doibase
  10.1103/RevModPhys.91.035001} {\bibfield  {journal} {\bibinfo  {journal}
  {Rev. Mod. Phys.}\ }\textbf {\bibinfo {volume} {91}},\ \bibinfo {pages}
  {035001} (\bibinfo {year} {2019})}\BibitemShut {NoStop}%
\bibitem [{\citenamefont {Julienne}(2012)}]{Julienne2012}%
  \BibitemOpen
  \bibfield  {author} {\bibinfo {author} {\bibfnamefont {P.~S.}\ \bibnamefont
  {Julienne}},\ }\href {\doibase 10.1038/nphys2386} {\bibfield  {journal}
  {\bibinfo  {journal} {Nature Physics}\ }\textbf {\bibinfo {volume} {8}},\
  \bibinfo {pages} {642} (\bibinfo {year} {2012})}\BibitemShut {NoStop}%
\bibitem [{\citenamefont {Puri}\ \emph {et~al.}(2019)\citenamefont {Puri},
  \citenamefont {Mills}, \citenamefont {Simbotin}, \citenamefont {Montgomery},
  \citenamefont {C{\^o}t{\'e}}, \citenamefont {Schneider}, \citenamefont
  {Suits},\ and\ \citenamefont {Hudson}}]{Hudson2019}%
  \BibitemOpen
  \bibfield  {author} {\bibinfo {author} {\bibfnamefont {P.}~\bibnamefont
  {Puri}}, \bibinfo {author} {\bibfnamefont {M.}~\bibnamefont {Mills}},
  \bibinfo {author} {\bibfnamefont {I.}~\bibnamefont {Simbotin}}, \bibinfo
  {author} {\bibfnamefont {J.~A.}\ \bibnamefont {Montgomery}}, \bibinfo
  {author} {\bibfnamefont {R.}~\bibnamefont {C{\^o}t{\'e}}}, \bibinfo {author}
  {\bibfnamefont {C.}~\bibnamefont {Schneider}}, \bibinfo {author}
  {\bibfnamefont {A.~G.}\ \bibnamefont {Suits}}, \ and\ \bibinfo {author}
  {\bibfnamefont {E.~R.}\ \bibnamefont {Hudson}},\ }\href {\doibase
  10.1038/s41557-019-0264-3} {\bibfield  {journal} {\bibinfo  {journal} {Nature
  Chemistry}\ }\textbf {\bibinfo {volume} {11}},\ \bibinfo {pages} {615}
  (\bibinfo {year} {2019})}\BibitemShut {NoStop}%
\bibitem [{\citenamefont {Weidinger}\ and\ \citenamefont
  {Gruebele}(2008)}]{QC1}%
  \BibitemOpen
  \bibfield  {author} {\bibinfo {author} {\bibfnamefont {D.}~\bibnamefont
  {Weidinger}}\ and\ \bibinfo {author} {\bibfnamefont {M.}~\bibnamefont
  {Gruebele}},\ }\href {\doibase
  https://doi.org/10.1016/j.chemphys.2008.01.059} {\bibfield  {journal}
  {\bibinfo  {journal} {Chemical Physics}\ }\textbf {\bibinfo {volume} {350}},\
  \bibinfo {pages} {139} (\bibinfo {year} {2008})},\ \bibinfo {note}
  {femtochemistry and Femtobiology}\BibitemShut {NoStop}%
\bibitem [{\citenamefont {Najafian}\ \emph
  {et~al.}(2020{\natexlab{a}})\citenamefont {Najafian}, \citenamefont {Meir},\
  and\ \citenamefont {Willitsch}}]{QC2}%
  \BibitemOpen
  \bibfield  {author} {\bibinfo {author} {\bibfnamefont {K.}~\bibnamefont
  {Najafian}}, \bibinfo {author} {\bibfnamefont {Z.}~\bibnamefont {Meir}}, \
  and\ \bibinfo {author} {\bibfnamefont {S.}~\bibnamefont {Willitsch}},\ }\href
  {\doibase 10.1039/D0CP03906C} {\bibfield  {journal} {\bibinfo  {journal}
  {Phys. Chem. Chem. Phys.}\ }\textbf {\bibinfo {volume} {22}},\ \bibinfo
  {pages} {23083} (\bibinfo {year} {2020}{\natexlab{a}})}\BibitemShut {NoStop}%
\bibitem [{\citenamefont {Häffner}\ \emph {et~al.}(2008)\citenamefont
  {Häffner}, \citenamefont {Roos},\ and\ \citenamefont {Blatt}}]{QC3}%
  \BibitemOpen
  \bibfield  {author} {\bibinfo {author} {\bibfnamefont {H.}~\bibnamefont
  {Häffner}}, \bibinfo {author} {\bibfnamefont {C.}~\bibnamefont {Roos}}, \
  and\ \bibinfo {author} {\bibfnamefont {R.}~\bibnamefont {Blatt}},\ }\href
  {\doibase https://doi.org/10.1016/j.physrep.2008.09.003} {\bibfield
  {journal} {\bibinfo  {journal} {Physics Reports}\ }\textbf {\bibinfo {volume}
  {469}},\ \bibinfo {pages} {155} (\bibinfo {year} {2008})}\BibitemShut
  {NoStop}%
\bibitem [{\citenamefont {Knight}\ \emph {et~al.}(2003)\citenamefont {Knight},
  \citenamefont {Hinds}, \citenamefont {Plenio}, \citenamefont {Wineland},
  \citenamefont {Barrett}, \citenamefont {Britton}, \citenamefont {Chiaverini},
  \citenamefont {DeMarco}, \citenamefont {Itano}, \citenamefont {Jelenković},
  \citenamefont {Langer}, \citenamefont {Leibfried}, \citenamefont {Meyer},
  \citenamefont {Rosenband},\ and\ \citenamefont {Schätz}}]{QC4}%
  \BibitemOpen
  \bibfield  {author} {\bibinfo {author} {\bibfnamefont {P.~L.}\ \bibnamefont
  {Knight}}, \bibinfo {author} {\bibfnamefont {E.~A.}\ \bibnamefont {Hinds}},
  \bibinfo {author} {\bibfnamefont {M.~B.}\ \bibnamefont {Plenio}}, \bibinfo
  {author} {\bibfnamefont {D.~J.}\ \bibnamefont {Wineland}}, \bibinfo {author}
  {\bibfnamefont {M.}~\bibnamefont {Barrett}}, \bibinfo {author} {\bibfnamefont
  {J.}~\bibnamefont {Britton}}, \bibinfo {author} {\bibfnamefont
  {J.}~\bibnamefont {Chiaverini}}, \bibinfo {author} {\bibfnamefont
  {B.}~\bibnamefont {DeMarco}}, \bibinfo {author} {\bibfnamefont {W.~M.}\
  \bibnamefont {Itano}}, \bibinfo {author} {\bibfnamefont {B.}~\bibnamefont
  {Jelenković}}, \bibinfo {author} {\bibfnamefont {C.}~\bibnamefont {Langer}},
  \bibinfo {author} {\bibfnamefont {D.}~\bibnamefont {Leibfried}}, \bibinfo
  {author} {\bibfnamefont {V.}~\bibnamefont {Meyer}}, \bibinfo {author}
  {\bibfnamefont {T.}~\bibnamefont {Rosenband}}, \ and\ \bibinfo {author}
  {\bibfnamefont {T.}~\bibnamefont {Schätz}},\ }\href {\doibase
  10.1098/rsta.2003.1205} {\bibfield  {journal} {\bibinfo  {journal}
  {Philosophical Transactions of the Royal Society of London. Series A:
  Mathematical, Physical and Engineering Sciences}\ }\textbf {\bibinfo {volume}
  {361}},\ \bibinfo {pages} {1349} (\bibinfo {year} {2003})}\BibitemShut
  {NoStop}%
\bibitem [{\citenamefont {Bruzewicz}\ \emph {et~al.}(2019)\citenamefont
  {Bruzewicz}, \citenamefont {Chiaverini}, \citenamefont {McConnell},\ and\
  \citenamefont {Sage}}]{QC5}%
  \BibitemOpen
  \bibfield  {author} {\bibinfo {author} {\bibfnamefont {C.~D.}\ \bibnamefont
  {Bruzewicz}}, \bibinfo {author} {\bibfnamefont {J.}~\bibnamefont
  {Chiaverini}}, \bibinfo {author} {\bibfnamefont {R.}~\bibnamefont
  {McConnell}}, \ and\ \bibinfo {author} {\bibfnamefont {J.~M.}\ \bibnamefont
  {Sage}},\ }\href {\doibase 10.1063/1.5088164} {\bibfield  {journal} {\bibinfo
   {journal} {Applied Physics Reviews}\ }\textbf {\bibinfo {volume} {6}},\
  \bibinfo {pages} {021314} (\bibinfo {year} {2019})},\ \Eprint
  {http://arxiv.org/abs/https://doi.org/10.1063/1.5088164}
  {https://doi.org/10.1063/1.5088164} \BibitemShut {NoStop}%
\bibitem [{\citenamefont {Lekitsch}\ \emph {et~al.}(2017)\citenamefont
  {Lekitsch}, \citenamefont {Weidt}, \citenamefont {Fowler}, \citenamefont
  {M{\o}lmer}, \citenamefont {Devitt}, \citenamefont {Wunderlich},\ and\
  \citenamefont {Hensinger}}]{QC6}%
  \BibitemOpen
  \bibfield  {author} {\bibinfo {author} {\bibfnamefont {B.}~\bibnamefont
  {Lekitsch}}, \bibinfo {author} {\bibfnamefont {S.}~\bibnamefont {Weidt}},
  \bibinfo {author} {\bibfnamefont {A.~G.}\ \bibnamefont {Fowler}}, \bibinfo
  {author} {\bibfnamefont {K.}~\bibnamefont {M{\o}lmer}}, \bibinfo {author}
  {\bibfnamefont {S.~J.}\ \bibnamefont {Devitt}}, \bibinfo {author}
  {\bibfnamefont {C.}~\bibnamefont {Wunderlich}}, \ and\ \bibinfo {author}
  {\bibfnamefont {W.~K.}\ \bibnamefont {Hensinger}},\ }\href {\doibase
  10.1126/sciadv.1601540} {\bibfield  {journal} {\bibinfo  {journal} {Science
  Advances}\ }\textbf {\bibinfo {volume} {3}} (\bibinfo {year} {2017}),\
  10.1126/sciadv.1601540}\BibitemShut {NoStop}%
\bibitem [{\citenamefont {Schindler}\ \emph {et~al.}(2013)\citenamefont
  {Schindler}, \citenamefont {Nigg}, \citenamefont {Monz}, \citenamefont
  {Barreiro}, \citenamefont {Martinez}, \citenamefont {Wang}, \citenamefont
  {Quint}, \citenamefont {Brandl}, \citenamefont {Nebendahl}, \citenamefont
  {Roos}, \citenamefont {Chwalla}, \citenamefont {Hennrich},\ and\
  \citenamefont {Blatt}}]{QC7}%
  \BibitemOpen
  \bibfield  {author} {\bibinfo {author} {\bibfnamefont {P.}~\bibnamefont
  {Schindler}}, \bibinfo {author} {\bibfnamefont {D.}~\bibnamefont {Nigg}},
  \bibinfo {author} {\bibfnamefont {T.}~\bibnamefont {Monz}}, \bibinfo {author}
  {\bibfnamefont {J.~T.}\ \bibnamefont {Barreiro}}, \bibinfo {author}
  {\bibfnamefont {E.}~\bibnamefont {Martinez}}, \bibinfo {author}
  {\bibfnamefont {S.~X.}\ \bibnamefont {Wang}}, \bibinfo {author}
  {\bibfnamefont {S.}~\bibnamefont {Quint}}, \bibinfo {author} {\bibfnamefont
  {M.~F.}\ \bibnamefont {Brandl}}, \bibinfo {author} {\bibfnamefont
  {V.}~\bibnamefont {Nebendahl}}, \bibinfo {author} {\bibfnamefont {C.~F.}\
  \bibnamefont {Roos}}, \bibinfo {author} {\bibfnamefont {M.}~\bibnamefont
  {Chwalla}}, \bibinfo {author} {\bibfnamefont {M.}~\bibnamefont {Hennrich}}, \
  and\ \bibinfo {author} {\bibfnamefont {R.}~\bibnamefont {Blatt}},\ }\href
  {\doibase 10.1088/1367-2630/15/12/123012} {\bibfield  {journal} {\bibinfo
  {journal} {New Journal of Physics}\ }\textbf {\bibinfo {volume} {15}},\
  \bibinfo {pages} {123012} (\bibinfo {year} {2013})}\BibitemShut {NoStop}%
\bibitem [{\citenamefont {Cirac}\ and\ \citenamefont {Zoller}(1995)}]{QC8}%
  \BibitemOpen
  \bibfield  {author} {\bibinfo {author} {\bibfnamefont {J.~I.}\ \bibnamefont
  {Cirac}}\ and\ \bibinfo {author} {\bibfnamefont {P.}~\bibnamefont {Zoller}},\
  }\href {\doibase 10.1103/PhysRevLett.74.4091} {\bibfield  {journal} {\bibinfo
   {journal} {Phys. Rev. Lett.}\ }\textbf {\bibinfo {volume} {74}},\ \bibinfo
  {pages} {4091} (\bibinfo {year} {1995})}\BibitemShut {NoStop}%
\bibitem [{\citenamefont {Mur-Petit}\ \emph {et~al.}(2012)\citenamefont
  {Mur-Petit}, \citenamefont {Garc\'{\i}a-Ripoll}, \citenamefont
  {P\'erez-R\'{\i}os}, \citenamefont {Campos-Mart\'{\i}nez}, \citenamefont
  {Hern\'andez},\ and\ \citenamefont {Willitsch}}]{Mur2012}%
  \BibitemOpen
  \bibfield  {author} {\bibinfo {author} {\bibfnamefont {J.}~\bibnamefont
  {Mur-Petit}}, \bibinfo {author} {\bibfnamefont {J.~J.}\ \bibnamefont
  {Garc\'{\i}a-Ripoll}}, \bibinfo {author} {\bibfnamefont {J.}~\bibnamefont
  {P\'erez-R\'{\i}os}}, \bibinfo {author} {\bibfnamefont {J.}~\bibnamefont
  {Campos-Mart\'{\i}nez}}, \bibinfo {author} {\bibfnamefont {M.~I.}\
  \bibnamefont {Hern\'andez}}, \ and\ \bibinfo {author} {\bibfnamefont
  {S.}~\bibnamefont {Willitsch}},\ }\href {\doibase 10.1103/PhysRevA.85.022308}
  {\bibfield  {journal} {\bibinfo  {journal} {Phys. Rev. A}\ }\textbf {\bibinfo
  {volume} {85}},\ \bibinfo {pages} {022308} (\bibinfo {year}
  {2012})}\BibitemShut {NoStop}%
\bibitem [{\citenamefont {Mur-Petit}\ \emph {et~al.}(2013)\citenamefont
  {Mur-Petit}, \citenamefont {P{\'e}rez-R{\'i}os}, \citenamefont
  {Campos-Mart{\'i}nez}, \citenamefont {Hern{\'a}ndez}, \citenamefont
  {Willitsch},\ and\ \citenamefont {Garc{\'i}a-Ripoll}}]{QLSchapter}%
  \BibitemOpen
  \bibfield  {author} {\bibinfo {author} {\bibfnamefont {J.}~\bibnamefont
  {Mur-Petit}}, \bibinfo {author} {\bibfnamefont {J.}~\bibnamefont
  {P{\'e}rez-R{\'i}os}}, \bibinfo {author} {\bibfnamefont {J.}~\bibnamefont
  {Campos-Mart{\'i}nez}}, \bibinfo {author} {\bibfnamefont {M.~I.}\
  \bibnamefont {Hern{\'a}ndez}}, \bibinfo {author} {\bibfnamefont
  {S.}~\bibnamefont {Willitsch}}, \ and\ \bibinfo {author} {\bibfnamefont
  {J.~J.}\ \bibnamefont {Garc{\'i}a-Ripoll}},\ }in\ \href@noop {} {\emph
  {\bibinfo {booktitle} {Architecture and Design of Molecule Logic Gates and
  Atom Circuits}}},\ \bibinfo {editor} {edited by\ \bibinfo {editor}
  {\bibfnamefont {N.}~\bibnamefont {Lorente}}\ and\ \bibinfo {editor}
  {\bibfnamefont {C.}~\bibnamefont {Joachim}}}\ (\bibinfo  {publisher}
  {Springer Berlin Heidelberg},\ \bibinfo {address} {Berlin, Heidelberg},\
  \bibinfo {year} {2013})\ pp.\ \bibinfo {pages} {267--277}\BibitemShut
  {NoStop}%
\bibitem [{\citenamefont {Wolf}\ \emph {et~al.}(2016)\citenamefont {Wolf},
  \citenamefont {Wan}, \citenamefont {Heip}, \citenamefont {Gebert},
  \citenamefont {Shi},\ and\ \citenamefont {Schmidt}}]{Wolf2016}%
  \BibitemOpen
  \bibfield  {author} {\bibinfo {author} {\bibfnamefont {F.}~\bibnamefont
  {Wolf}}, \bibinfo {author} {\bibfnamefont {Y.}~\bibnamefont {Wan}}, \bibinfo
  {author} {\bibfnamefont {J.~C.}\ \bibnamefont {Heip}}, \bibinfo {author}
  {\bibfnamefont {F.}~\bibnamefont {Gebert}}, \bibinfo {author} {\bibfnamefont
  {C.}~\bibnamefont {Shi}}, \ and\ \bibinfo {author} {\bibfnamefont {P.~O.}\
  \bibnamefont {Schmidt}},\ }\href {\doibase 10.1038/nature16513} {\bibfield
  {journal} {\bibinfo  {journal} {Nature}\ }\textbf {\bibinfo {volume} {530}},\
  \bibinfo {pages} {457} (\bibinfo {year} {2016})}\BibitemShut {NoStop}%
\bibitem [{\citenamefont {Sinhal}\ \emph {et~al.}(2020)\citenamefont {Sinhal},
  \citenamefont {Meir}, \citenamefont {Najafian}, \citenamefont {Hegi},\ and\
  \citenamefont {Willitsch}}]{Sinhal12020}%
  \BibitemOpen
  \bibfield  {author} {\bibinfo {author} {\bibfnamefont {M.}~\bibnamefont
  {Sinhal}}, \bibinfo {author} {\bibfnamefont {Z.}~\bibnamefont {Meir}},
  \bibinfo {author} {\bibfnamefont {K.}~\bibnamefont {Najafian}}, \bibinfo
  {author} {\bibfnamefont {G.}~\bibnamefont {Hegi}}, \ and\ \bibinfo {author}
  {\bibfnamefont {S.}~\bibnamefont {Willitsch}},\ }\href {\doibase
  10.1126/science.aaz9837} {\bibfield  {journal} {\bibinfo  {journal}
  {Science}\ }\textbf {\bibinfo {volume} {367}},\ \bibinfo {pages} {1213}
  (\bibinfo {year} {2020})}\BibitemShut {NoStop}%
\bibitem [{\citenamefont {Najafian}\ \emph
  {et~al.}(2020{\natexlab{b}})\citenamefont {Najafian}, \citenamefont {Meir},
  \citenamefont {Sinhal},\ and\ \citenamefont {Willitsch}}]{Najafian20202}%
  \BibitemOpen
  \bibfield  {author} {\bibinfo {author} {\bibfnamefont {K.}~\bibnamefont
  {Najafian}}, \bibinfo {author} {\bibfnamefont {Z.}~\bibnamefont {Meir}},
  \bibinfo {author} {\bibfnamefont {M.}~\bibnamefont {Sinhal}}, \ and\ \bibinfo
  {author} {\bibfnamefont {S.}~\bibnamefont {Willitsch}},\ }\href {\doibase
  10.1038/s41467-020-18170-9} {\bibfield  {journal} {\bibinfo  {journal}
  {Nature Communications}\ }\textbf {\bibinfo {volume} {11}},\ \bibinfo {pages}
  {4470} (\bibinfo {year} {2020}{\natexlab{b}})}\BibitemShut {NoStop}%
\bibitem [{\citenamefont {Collopy}\ \emph {et~al.}(2021)\citenamefont
  {Collopy}, \citenamefont {Leibrandt}, \citenamefont {Leibfried},\ and\
  \citenamefont {Chou}}]{Collopy2021}%
  \BibitemOpen
  \bibfield  {author} {\bibinfo {author} {\bibfnamefont {A.}~\bibnamefont
  {Collopy}}, \bibinfo {author} {\bibfnamefont {D.~R.}\ \bibnamefont
  {Leibrandt}}, \bibinfo {author} {\bibfnamefont {D.~G.}\ \bibnamefont
  {Leibfried}}, \ and\ \bibinfo {author} {\bibfnamefont {C.-W.}\ \bibnamefont
  {Chou}},\ }in\ \href {\doibase 10.1117/12.2586882} {\emph {\bibinfo
  {booktitle} {Optical and Quantum Sensing and Precision Metrology}}},\ Vol.\
  \bibinfo {volume} {11700},\ \bibinfo {editor} {edited by\ \bibinfo {editor}
  {\bibfnamefont {S.~M.}\ \bibnamefont {Shahriar}}\ and\ \bibinfo {editor}
  {\bibfnamefont {J.}~\bibnamefont {Scheuer}}},\ \bibinfo {organization}
  {International Society for Optics and Photonics}\ (\bibinfo  {publisher}
  {SPIE},\ \bibinfo {year} {2021})\BibitemShut {NoStop}%
\bibitem [{\citenamefont {Campbell}\ and\ \citenamefont
  {Hudson}(2020)}]{PhysRevLett.125.120501}%
  \BibitemOpen
  \bibfield  {author} {\bibinfo {author} {\bibfnamefont {W.~C.}\ \bibnamefont
  {Campbell}}\ and\ \bibinfo {author} {\bibfnamefont {E.~R.}\ \bibnamefont
  {Hudson}},\ }\href {\doibase 10.1103/PhysRevLett.125.120501} {\bibfield
  {journal} {\bibinfo  {journal} {Phys. Rev. Lett.}\ }\textbf {\bibinfo
  {volume} {125}},\ \bibinfo {pages} {120501} (\bibinfo {year}
  {2020})}\BibitemShut {NoStop}%
\bibitem [{\citenamefont {Kleinbach}\ \emph {et~al.}(2018)\citenamefont
  {Kleinbach}, \citenamefont {Engel}, \citenamefont {Dieterle}, \citenamefont
  {L{\"o}w}, \citenamefont {Pfau},\ and\ \citenamefont
  {Meinert}}]{Kleinbach_2018}%
  \BibitemOpen
  \bibfield  {author} {\bibinfo {author} {\bibfnamefont {K.~S.}\ \bibnamefont
  {Kleinbach}}, \bibinfo {author} {\bibfnamefont {F.}~\bibnamefont {Engel}},
  \bibinfo {author} {\bibfnamefont {T.}~\bibnamefont {Dieterle}}, \bibinfo
  {author} {\bibfnamefont {R.}~\bibnamefont {L{\"o}w}}, \bibinfo {author}
  {\bibfnamefont {T.}~\bibnamefont {Pfau}}, \ and\ \bibinfo {author}
  {\bibfnamefont {F.}~\bibnamefont {Meinert}},\ }\href {\doibase
  10.1103/PhysRevLett.120.193401} {\bibfield  {journal} {\bibinfo  {journal}
  {Phys. Rev. Lett.}\ }\textbf {\bibinfo {volume} {120}},\ \bibinfo {pages}
  {193401} (\bibinfo {year} {2018})}\BibitemShut {NoStop}%
\bibitem [{\citenamefont {Schurer}\ \emph {et~al.}(2014)\citenamefont
  {Schurer}, \citenamefont {Schmelcher},\ and\ \citenamefont
  {Negretti}}]{Meso1}%
  \BibitemOpen
  \bibfield  {author} {\bibinfo {author} {\bibfnamefont {J.~M.}\ \bibnamefont
  {Schurer}}, \bibinfo {author} {\bibfnamefont {P.}~\bibnamefont {Schmelcher}},
  \ and\ \bibinfo {author} {\bibfnamefont {A.}~\bibnamefont {Negretti}},\
  }\href@noop {} {\bibfield  {journal} {\bibinfo  {journal} {Phys. Rev. A}\
  }\textbf {\bibinfo {volume} {90}},\ \bibinfo {pages} {033601} (\bibinfo
  {year} {2014})}\BibitemShut {NoStop}%
\bibitem [{\citenamefont {Schurer}\ \emph {et~al.}(2017)\citenamefont
  {Schurer}, \citenamefont {Negretti},\ and\ \citenamefont
  {Schmelcher}}]{Meso2}%
  \BibitemOpen
  \bibfield  {author} {\bibinfo {author} {\bibfnamefont {J.~M.}\ \bibnamefont
  {Schurer}}, \bibinfo {author} {\bibfnamefont {A.}~\bibnamefont {Negretti}}, \
  and\ \bibinfo {author} {\bibfnamefont {P.}~\bibnamefont {Schmelcher}},\
  }\href@noop {} {\bibfield  {journal} {\bibinfo  {journal} {Phys. Rev. Lett.}\
  }\textbf {\bibinfo {volume} {119}},\ \bibinfo {pages} {063001} (\bibinfo
  {year} {2017})}\BibitemShut {NoStop}%
\bibitem [{\citenamefont {C\^ot\'e}\ \emph {et~al.}(2002)\citenamefont
  {C\^ot\'e}, \citenamefont {Kharchenko},\ and\ \citenamefont {Lukin}}]{Meso3}%
  \BibitemOpen
  \bibfield  {author} {\bibinfo {author} {\bibfnamefont {R.}~\bibnamefont
  {C\^ot\'e}}, \bibinfo {author} {\bibfnamefont {V.}~\bibnamefont
  {Kharchenko}}, \ and\ \bibinfo {author} {\bibfnamefont {M.~D.}\ \bibnamefont
  {Lukin}},\ }\href@noop {} {\bibfield  {journal} {\bibinfo  {journal} {Phys.
  Rev. Lett.}\ }\textbf {\bibinfo {volume} {89}},\ \bibinfo {pages} {093001}
  (\bibinfo {year} {2002})}\BibitemShut {NoStop}%
\bibitem [{\citenamefont {Astrakharchik}\ \emph {et~al.}(2020)\citenamefont
  {Astrakharchik}, \citenamefont {Ardila}, \citenamefont {Schmidt},
  \citenamefont {Jachymski},\ and\ \citenamefont
  {Negretti}}]{astrakharchik2020ionic}%
  \BibitemOpen
  \bibfield  {author} {\bibinfo {author} {\bibfnamefont {G.~E.}\ \bibnamefont
  {Astrakharchik}}, \bibinfo {author} {\bibfnamefont {L.~A.~P.}\ \bibnamefont
  {Ardila}}, \bibinfo {author} {\bibfnamefont {R.}~\bibnamefont {Schmidt}},
  \bibinfo {author} {\bibfnamefont {K.}~\bibnamefont {Jachymski}}, \ and\
  \bibinfo {author} {\bibfnamefont {A.}~\bibnamefont {Negretti}},\ }\href@noop
  {} {\enquote {\bibinfo {title} {Ionic polaron in a bose-einstein
  condensate},}\ } (\bibinfo {year} {2020}),\ \Eprint
  {http://arxiv.org/abs/2005.12033} {arXiv:2005.12033 [cond-mat.quant-gas]}
  \BibitemShut {NoStop}%
\bibitem [{\citenamefont {Massignan}\ \emph {et~al.}(2005)\citenamefont
  {Massignan}, \citenamefont {Pethick},\ and\ \citenamefont
  {Smith}}]{Massignan2005}%
  \BibitemOpen
  \bibfield  {author} {\bibinfo {author} {\bibfnamefont {P.}~\bibnamefont
  {Massignan}}, \bibinfo {author} {\bibfnamefont {C.~J.}\ \bibnamefont
  {Pethick}}, \ and\ \bibinfo {author} {\bibfnamefont {H.}~\bibnamefont
  {Smith}},\ }\href {\doibase 10.1103/PhysRevA.71.023606} {\bibfield  {journal}
  {\bibinfo  {journal} {Phys. Rev. A}\ }\textbf {\bibinfo {volume} {71}},\
  \bibinfo {pages} {023606} (\bibinfo {year} {2005})}\BibitemShut {NoStop}%
\bibitem [{\citenamefont {Dieterle}\ \emph {et~al.}(2020)\citenamefont
  {Dieterle}, \citenamefont {Berngruber}, \citenamefont {H\"olzl},
  \citenamefont {L\"ow}, \citenamefont {Jachymski}, \citenamefont {Pfau},\ and\
  \citenamefont {Meinert}}]{dieterle2020transport}%
  \BibitemOpen
  \bibfield  {author} {\bibinfo {author} {\bibfnamefont {T.}~\bibnamefont
  {Dieterle}}, \bibinfo {author} {\bibfnamefont {M.}~\bibnamefont
  {Berngruber}}, \bibinfo {author} {\bibfnamefont {C.}~\bibnamefont {H\"olzl}},
  \bibinfo {author} {\bibfnamefont {R.}~\bibnamefont {L\"ow}}, \bibinfo
  {author} {\bibfnamefont {K.}~\bibnamefont {Jachymski}}, \bibinfo {author}
  {\bibfnamefont {T.}~\bibnamefont {Pfau}}, \ and\ \bibinfo {author}
  {\bibfnamefont {F.}~\bibnamefont {Meinert}},\ }\href {\doibase
  10.1103/PhysRevA.102.041301} {\bibfield  {journal} {\bibinfo  {journal}
  {Phys. Rev. A}\ }\textbf {\bibinfo {volume} {102}},\ \bibinfo {pages}
  {041301} (\bibinfo {year} {2020})}\BibitemShut {NoStop}%
\bibitem [{\citenamefont {Midya}\ \emph {et~al.}(2016)\citenamefont {Midya},
  \citenamefont {Tomza}, \citenamefont {Schmidt},\ and\ \citenamefont
  {Lemeshko}}]{Lemeshko2016}%
  \BibitemOpen
  \bibfield  {author} {\bibinfo {author} {\bibfnamefont {B.}~\bibnamefont
  {Midya}}, \bibinfo {author} {\bibfnamefont {M.}~\bibnamefont {Tomza}},
  \bibinfo {author} {\bibfnamefont {R.}~\bibnamefont {Schmidt}}, \ and\
  \bibinfo {author} {\bibfnamefont {M.}~\bibnamefont {Lemeshko}},\ }\href
  {\doibase 10.1103/PhysRevA.94.041601} {\bibfield  {journal} {\bibinfo
  {journal} {Phys. Rev. A}\ }\textbf {\bibinfo {volume} {94}},\ \bibinfo
  {pages} {041601} (\bibinfo {year} {2016})}\BibitemShut {NoStop}%
\bibitem [{\citenamefont {Mukherjee}\ \emph {et~al.}(2015)\citenamefont
  {Mukherjee}, \citenamefont {Ates}, \citenamefont {Li},\ and\ \citenamefont
  {W\"uster}}]{Mukherjee2015}%
  \BibitemOpen
  \bibfield  {author} {\bibinfo {author} {\bibfnamefont {R.}~\bibnamefont
  {Mukherjee}}, \bibinfo {author} {\bibfnamefont {C.}~\bibnamefont {Ates}},
  \bibinfo {author} {\bibfnamefont {W.}~\bibnamefont {Li}}, \ and\ \bibinfo
  {author} {\bibfnamefont {S.}~\bibnamefont {W\"uster}},\ }\href {\doibase
  10.1103/PhysRevLett.115.040401} {\bibfield  {journal} {\bibinfo  {journal}
  {Phys. Rev. Lett.}\ }\textbf {\bibinfo {volume} {115}},\ \bibinfo {pages}
  {040401} (\bibinfo {year} {2015})}\BibitemShut {NoStop}%
\bibitem [{\citenamefont {Kr\"ukow}\ \emph {et~al.}(2016)\citenamefont
  {Kr\"ukow}, \citenamefont {Mohammadi}, \citenamefont {H\"arter},\ and\
  \citenamefont {Hecker~Denschlag}}]{Artjom2016}%
  \BibitemOpen
  \bibfield  {author} {\bibinfo {author} {\bibfnamefont {A.}~\bibnamefont
  {Kr\"ukow}}, \bibinfo {author} {\bibfnamefont {A.}~\bibnamefont {Mohammadi}},
  \bibinfo {author} {\bibfnamefont {A.}~\bibnamefont {H\"arter}}, \ and\
  \bibinfo {author} {\bibfnamefont {J.}~\bibnamefont {Hecker~Denschlag}},\
  }\href {\doibase 10.1103/PhysRevA.94.030701} {\bibfield  {journal} {\bibinfo
  {journal} {Phys. Rev. A}\ }\textbf {\bibinfo {volume} {94}},\ \bibinfo
  {pages} {030701} (\bibinfo {year} {2016})}\BibitemShut {NoStop}%
\bibitem [{\citenamefont {Pérez-Ríos}(2021)}]{JPR2021}%
  \BibitemOpen
  \bibfield  {author} {\bibinfo {author} {\bibfnamefont {J.}~\bibnamefont
  {Pérez-Ríos}},\ }\href {\doibase 10.1080/00268976.2021.1881637} {\bibfield
  {journal} {\bibinfo  {journal} {Molecular Physics}\ }\textbf {\bibinfo
  {volume} {119}},\ \bibinfo {pages} {e1881637} (\bibinfo {year} {2021})},\
  \Eprint {http://arxiv.org/abs/https://doi.org/10.1080/00268976.2021.1881637}
  {https://doi.org/10.1080/00268976.2021.1881637} \BibitemShut {NoStop}%
\bibitem [{\citenamefont {Mohammadi}\ \emph {et~al.}(2021)\citenamefont
  {Mohammadi}, \citenamefont {Kr\"ukow}, \citenamefont {Mahdian}, \citenamefont
  {Dei\ss{}}, \citenamefont {P\'erez-R\'{\i}os}, \citenamefont {da~Silva},
  \citenamefont {Raoult}, \citenamefont {Dulieu},\ and\ \citenamefont
  {Hecker~Denschlag}}]{Amir2021}%
  \BibitemOpen
  \bibfield  {author} {\bibinfo {author} {\bibfnamefont {A.}~\bibnamefont
  {Mohammadi}}, \bibinfo {author} {\bibfnamefont {A.}~\bibnamefont {Kr\"ukow}},
  \bibinfo {author} {\bibfnamefont {A.}~\bibnamefont {Mahdian}}, \bibinfo
  {author} {\bibfnamefont {M.}~\bibnamefont {Dei\ss{}}}, \bibinfo {author}
  {\bibfnamefont {J.}~\bibnamefont {P\'erez-R\'{\i}os}}, \bibinfo {author}
  {\bibfnamefont {H.}~\bibnamefont {da~Silva}}, \bibinfo {author}
  {\bibfnamefont {M.}~\bibnamefont {Raoult}}, \bibinfo {author} {\bibfnamefont
  {O.}~\bibnamefont {Dulieu}}, \ and\ \bibinfo {author} {\bibfnamefont
  {J.}~\bibnamefont {Hecker~Denschlag}},\ }\href {\doibase
  10.1103/PhysRevResearch.3.013196} {\bibfield  {journal} {\bibinfo  {journal}
  {Phys. Rev. Research}\ }\textbf {\bibinfo {volume} {3}},\ \bibinfo {pages}
  {013196} (\bibinfo {year} {2021})}\BibitemShut {NoStop}%
\bibitem [{\citenamefont {P\'erez-R\'{\i}os}(2019)}]{JPR2019}%
  \BibitemOpen
  \bibfield  {author} {\bibinfo {author} {\bibfnamefont {J.}~\bibnamefont
  {P\'erez-R\'{\i}os}},\ }\href {\doibase 10.1103/PhysRevA.99.022707}
  {\bibfield  {journal} {\bibinfo  {journal} {Phys. Rev. A}\ }\textbf {\bibinfo
  {volume} {99}},\ \bibinfo {pages} {022707} (\bibinfo {year}
  {2019})}\BibitemShut {NoStop}%
\bibitem [{\citenamefont {Jachymski}\ and\ \citenamefont
  {Meinert}(2020)}]{Jachymski2020}%
  \BibitemOpen
  \bibfield  {author} {\bibinfo {author} {\bibfnamefont {K.}~\bibnamefont
  {Jachymski}}\ and\ \bibinfo {author} {\bibfnamefont {F.}~\bibnamefont
  {Meinert}},\ }\href {\doibase 10.3390/app10072371} {\bibfield  {journal}
  {\bibinfo  {journal} {Applied Sciences}\ }\textbf {\bibinfo {volume} {10}}
  (\bibinfo {year} {2020}),\ 10.3390/app10072371}\BibitemShut {NoStop}%
\bibitem [{\citenamefont {Hirzler}\ \emph {et~al.}(2020)\citenamefont
  {Hirzler}, \citenamefont {Trimby}, \citenamefont {Lous}, \citenamefont
  {Groenenboom}, \citenamefont {Gerritsma},\ and\ \citenamefont
  {P\'erez-R\'{\i}os}}]{Hirzler2020}%
  \BibitemOpen
  \bibfield  {author} {\bibinfo {author} {\bibfnamefont {H.}~\bibnamefont
  {Hirzler}}, \bibinfo {author} {\bibfnamefont {E.}~\bibnamefont {Trimby}},
  \bibinfo {author} {\bibfnamefont {R.~S.}\ \bibnamefont {Lous}}, \bibinfo
  {author} {\bibfnamefont {G.~C.}\ \bibnamefont {Groenenboom}}, \bibinfo
  {author} {\bibfnamefont {R.}~\bibnamefont {Gerritsma}}, \ and\ \bibinfo
  {author} {\bibfnamefont {J.}~\bibnamefont {P\'erez-R\'{\i}os}},\ }\href
  {\doibase 10.1103/PhysRevResearch.2.033232} {\bibfield  {journal} {\bibinfo
  {journal} {Phys. Rev. Research}\ }\textbf {\bibinfo {volume} {2}},\ \bibinfo
  {pages} {033232} (\bibinfo {year} {2020})}\BibitemShut {NoStop}%
\bibitem [{\citenamefont {Hoang}\ \emph {et~al.}(2020)\citenamefont {Hoang},
  \citenamefont {Jau}, \citenamefont {Overstreet},\ and\ \citenamefont
  {Schwindt}}]{YbH+}%
  \BibitemOpen
  \bibfield  {author} {\bibinfo {author} {\bibfnamefont {T.~M.}\ \bibnamefont
  {Hoang}}, \bibinfo {author} {\bibfnamefont {Y.-Y.}\ \bibnamefont {Jau}},
  \bibinfo {author} {\bibfnamefont {R.}~\bibnamefont {Overstreet}}, \ and\
  \bibinfo {author} {\bibfnamefont {P.~D.~D.}\ \bibnamefont {Schwindt}},\
  }\href {\doibase 10.1103/PhysRevA.101.022705} {\bibfield  {journal} {\bibinfo
   {journal} {Phys. Rev. A}\ }\textbf {\bibinfo {volume} {101}},\ \bibinfo
  {pages} {022705} (\bibinfo {year} {2020})}\BibitemShut {NoStop}%
\bibitem [{\citenamefont {Sugiyama}\ and\ \citenamefont {Yoda}(1997)}]{YbH+2}%
  \BibitemOpen
  \bibfield  {author} {\bibinfo {author} {\bibfnamefont {K.}~\bibnamefont
  {Sugiyama}}\ and\ \bibinfo {author} {\bibfnamefont {J.}~\bibnamefont
  {Yoda}},\ }\href {\doibase 10.1103/PhysRevA.55.R10} {\bibfield  {journal}
  {\bibinfo  {journal} {Phys. Rev. A}\ }\textbf {\bibinfo {volume} {55}},\
  \bibinfo {pages} {R10} (\bibinfo {year} {1997})}\BibitemShut {NoStop}%
\bibitem [{\citenamefont {Sugiyama}\ and\ \citenamefont {Yoda}(1995)}]{YbH+3}%
  \BibitemOpen
  \bibfield  {author} {\bibinfo {author} {\bibfnamefont {K.}~\bibnamefont
  {Sugiyama}}\ and\ \bibinfo {author} {\bibfnamefont {J.}~\bibnamefont
  {Yoda}},\ }\href {\doibase 10.1143/jjap.34.l584} {\bibfield  {journal}
  {\bibinfo  {journal} {Japanese Journal of Applied Physics}\ }\textbf
  {\bibinfo {volume} {34}},\ \bibinfo {pages} {L584} (\bibinfo {year}
  {1995})}\BibitemShut {NoStop}%
\bibitem [{\citenamefont {Lamb}\ \emph {et~al.}(2012)\citenamefont {Lamb},
  \citenamefont {McCann}, \citenamefont {McLaughlin}, \citenamefont {Goold},
  \citenamefont {Wells},\ and\ \citenamefont {Lane}}]{PotYbRbplus}%
  \BibitemOpen
  \bibfield  {author} {\bibinfo {author} {\bibfnamefont {H.~D.~L.}\
  \bibnamefont {Lamb}}, \bibinfo {author} {\bibfnamefont {J.~F.}\ \bibnamefont
  {McCann}}, \bibinfo {author} {\bibfnamefont {B.~M.}\ \bibnamefont
  {McLaughlin}}, \bibinfo {author} {\bibfnamefont {J.}~\bibnamefont {Goold}},
  \bibinfo {author} {\bibfnamefont {N.}~\bibnamefont {Wells}}, \ and\ \bibinfo
  {author} {\bibfnamefont {I.}~\bibnamefont {Lane}},\ }\href {\doibase
  10.1103/PhysRevA.86.022716} {\bibfield  {journal} {\bibinfo  {journal} {Phys.
  Rev. A}\ }\textbf {\bibinfo {volume} {86}},\ \bibinfo {pages} {022716}
  (\bibinfo {year} {2012})}\BibitemShut {NoStop}%
\bibitem [{\citenamefont {Lemeshko}\ and\ \citenamefont
  {Friedrich}(2009)}]{Bretislav}%
  \BibitemOpen
  \bibfield  {author} {\bibinfo {author} {\bibfnamefont {M.}~\bibnamefont
  {Lemeshko}}\ and\ \bibinfo {author} {\bibfnamefont {B.}~\bibnamefont
  {Friedrich}},\ }\href {\doibase 10.1103/PhysRevLett.103.053003} {\bibfield
  {journal} {\bibinfo  {journal} {Phys. Rev. Lett.}\ }\textbf {\bibinfo
  {volume} {103}},\ \bibinfo {pages} {053003} (\bibinfo {year}
  {2009})}\BibitemShut {NoStop}%
\bibitem [{\citenamefont {Tomza}\ \emph {et~al.}(2015)\citenamefont {Tomza},
  \citenamefont {Koch},\ and\ \citenamefont {Moszynski}}]{PotYbLiplus}%
  \BibitemOpen
  \bibfield  {author} {\bibinfo {author} {\bibfnamefont {M.}~\bibnamefont
  {Tomza}}, \bibinfo {author} {\bibfnamefont {C.~P.}\ \bibnamefont {Koch}}, \
  and\ \bibinfo {author} {\bibfnamefont {R.}~\bibnamefont {Moszynski}},\ }\href
  {\doibase 10.1103/PhysRevA.91.042706} {\bibfield  {journal} {\bibinfo
  {journal} {Phys. Rev. A}\ }\textbf {\bibinfo {volume} {91}},\ \bibinfo
  {pages} {042706} (\bibinfo {year} {2015})}\BibitemShut {NoStop}%
\bibitem [{\citenamefont {Pérez-Ríos}\ and\ \citenamefont
  {Greene}(2015)}]{JPR2015}%
  \BibitemOpen
  \bibfield  {author} {\bibinfo {author} {\bibfnamefont {J.}~\bibnamefont
  {Pérez-Ríos}}\ and\ \bibinfo {author} {\bibfnamefont {C.~H.}\ \bibnamefont
  {Greene}},\ }\href@noop {} {\bibfield  {journal} {\bibinfo  {journal} {The
  Journal of Chemical Physics}\ }\textbf {\bibinfo {volume} {143}},\ \bibinfo
  {pages} {041105} (\bibinfo {year} {2015})}\BibitemShut {NoStop}%
\end{thebibliography}

%

\end{document}